\documentclass[aps,preprint,showpacs,preprintnumbers,amsmath,amssymb]{revtex4}

\begin{document}

%\preprint{APS/123-QED}

\title{Self-phase modulation of
\\ spherical gravitational waves}

\author{J.T.\ Mendon\c{c}a}
\author{V.\ Cardoso}
 \email{titomend@ist.utl.pt}
\affiliation{GoLP and CENTRA, Instituto Superior T\'{e}cnico,
1049-001 Lisboa, Portugal}

\author{M.\ Marklund}
\affiliation{Department of Electromagnetics, Chalmers University of
  Technology, SE-412 96 G\"oteborg, Sweden}

\author{M.\ Servin}
\author{G.\ Brodin}
\affiliation{Department of Plasma Physics, SE-901 87 Ume{\aa}, Sweden} 

\date{May 4, 2004}

\begin{abstract}

Self-phase modulation of spherical gravitational wavepackets
propagating in a flat space-time in the presence of a tenuous
distribution of matter is considered. Analogies with 
respect to similar effects in nonlinear 
optics are explored.
Self phase modulation of waves emitted from a single source
 can eventually lead to an efficient
energy dilution of the gravitational wave energy over an
increasingly large spectral range. An explicit criterium 
for the occurrence of a significant spectral energy dilution
is established.

\end{abstract}

\pacs{04.30.Nk, 42.65.-k, 95.30.Sf}

\maketitle

\section{Introduction}

It is well known that the Einstein's equation describing 
gravitational waves is strongly nonlinear
\cite{weinberg,landau}. Nonlinear wave processes 
similar to those observed in Optics can then eventually occur.
Recently, the possible occurrence of self-phase modulation, harmonic 
generation and nonlinear wave mixing was considered \cite{mend02,servin}.

This is an important issue in two different aspects. First, from a
theoretical point of view 
it is important to explore and to understand the similarities and differences
between gravitational wave phenomena and effects in Nonlinear
Optics that have been 
tested in laboratory. Second, and in more practical terms, this is important
in what concerns the possible detection of gravitational waves.
Detectors have been designed and built under the assumption that the
frequency spectrum  
of gravitational waves emitted by astronomical objects is conserved, and that 
the wave intensities decrease as the inverse of the square of the distance. 
However, the nonlinear processes can eventually lead to spectral energy 
dilution such that the energy density received within the
detectable frequency 
bandwidth is significantly decreased. Other processes eventually contributing
to energy dilution could be due to the coupling with plasma waves
\cite{brodin}  
and with photons \cite{mend01,mosquera}.  

For waves emitted from a single astronomical source, the main nonlinear effects
that can occur are self-phase modulation (for short pulses, 
of the order of a few cycles) and harmonic cascades (for longer pulses). 
Attention was however called to the fact that, for parallel propagation, 
the strong nonlinearities associated with empty and flat space-time
exactly cancel each other \cite{servin}. The existence of such a negative result is
apparently due to the absence of gravitational wave dispersion. 
Note, however, that anti-parallel wave configurations lead to a
nonlinear coupling \cite{faraoni}, but clearly these are not relevant to 
waves emitted by single sources.

In this work, we return to the problem of parallel wave interactions. 
We will focus on self-phase modulation of spherical waves emitted 
by isolated sources.
For simplicity, we will consider a flat space-time, filled with a
tenuous distribution of matter \cite{tenuous}. The matter distribution will
guarantee the 
existence of wave dispersion, which is an important ingredient of
self-phase modulation in  
Nonlinear Optics \cite{luis,alfano}. The nonlinear wave equation is
established in  
Section II, and the dispersion properties of linear waves is discussed
in Section III. 
Nonlinear evolution of a spherical gravitational wavepacket is studied
in Section IV, 
where a necessary criterium for the occurrence of a significant amount
of self-phase modulation  
is established. Finally, in Section V we state the conclusions.

\section{Nonlinear wave equation}

We consider propagation of small amplitude gravitational waves, in a
region of space-time where we have a tenuous distribution of matter \cite{tenuous}. 
We can then, in a first approximation, neglect the background field
curvature.

We are considering a flat space-time, perturbed by a small
amplitude gravitational wave. This can be described by the metric
tensor elements 

\begin{equation}
g_{ij} = \eta_{ij} + h_{ij}, \label{e:2.1} \end{equation}
 where $|h_{ij}| \ll 1$ represent the gravitational wave, and
 $\eta_{ij}$ are the  
metric tensor elements of flat space-time

\begin{equation}
\eta_{00} = 1 \quad , \quad \eta_{ii} = -1 \quad (i = 1, 2, 3)
\quad , \quad  \eta_{ij} = 0 \quad (i \neq j). \label{eq:2.2}
\end{equation}

In this case, we can derive from Einstein's equation  the 
following nonlinear wave equation

\begin{equation}
\Box^2 h_{ik} = - 2 \kappa S_{ij} + 2 R_{ik}^{(3)}. 
\label{eq:2.3} \end{equation}
where $\kappa = (8 \pi / c^2) G$, and $G$ is the gravitational constant,
and $\Box^2$ is the d'Alembert operator in the usual form

\begin{equation}
\Box^2 = \partial^j \partial_j = 
\eta^{ij} \partial_i \partial_j. \label{eq:2.4}
\end{equation}
where we use $\partial_i \equiv \partial / \partial x^i$.
The nonlinear term $R_{ik}^{(3)}$ contains third order nonlinearities
of the Ricci tensor. It can be explicitly written as

\begin{eqnarray}
R_{ik}^{(3)} = - \frac{1}{4} \eta^{nm} h^{lp} \left[ \left(
\partial_j h_{mi} + \partial_i
h^{ml} -\partial_m h_{il}\right) 
\left( \partial_n h_{pk}
+\partial_k h_{pn} - \partial_p
h_{kn} \right) \right. 
\nonumber \\ - \left. \left(
\partial_k h_{pi} + \partial_i
h_{pk} - \partial_p h_{ik}
\right) \partial_l h_{np} \right].
\label{eq:2.5} \end{eqnarray}

Here we have neglected the second-order nonlinear term,
$R_{ik}^{(2)} = 0$. Note that the second order part of
the Ricci tensor does not vanish identically unless the response due to
the pseudo energy-momentum tensor is taken into account \cite{servin}. Combined with
the original perturbation, these second order terms contribute to the
self phase modulation of the wave. The equation studied here may
therefore be regarded as a model equation for the nonlinear dynamics
of gravitational waves. 
In equation
(\ref{eq:2.3}) we have also included the linear dispersion term associated with
the matter distribution $ \kappa S_{ij}$, where $S_{ij}$ is related 
with the energy-momentum tensor $T_{ij}$ by

\begin{equation}
S_{ij} = T_{ij} - \frac{1}{2} \eta_{ij} T \label{eq:2.6}
\end{equation}
where $T = T_i^i$ is the trace.

\section{Linear dispersion relation}

Let us first consider the properties of a linear wave, propagating
radially from a given 
point source. In order to discuss the dispersion properties of this
wave we first  
consider the linearized wave equation

\begin{equation}
\Box^2 h_{ik} = - \kappa S_{ij}, 
\label{eq:3.1} \end{equation}
and assume a plane wave solution of the form

\begin{equation}
h_{ij} = \epsilon_{ij} A \exp[i q_n x^n],
\label{eq:3.2} \end{equation} 
where $i = \sqrt{-1}$, $A$ is the amplitude,
$\epsilon_{ij}$ is unit polarization tensor such that
$\epsilon_{ij}^* \epsilon_{ij}=1$, and $q_n$ are the components
of the four-wavevector.
If the scale of variation of the amplitude $A$ is much larger than
the typical wavelength, we can use $\partial_j h_{ik} = i q_j h_{ik}$, 
and write the linear wave equation as

\begin{equation}
\eta^{jn} q_j q_n h_{ik} =  2 \kappa S_{ik}. 
\label{eq:3.3} \end{equation}

The perturbed energy-momentum tensor can be considered proportional to the
local amplitude of the gravitational wave: $S_{ik} = w_{ik} A$, where
the tensor  
$w_{ik}$ depends on the properties of the medium. 
We are then lead to the following linear dispersion relation

\begin{equation}
\eta^{jn} q_j q_n = w, \label{eq:3.4} \end{equation}
where we have used $w = 2 \kappa \epsilon^{ik*} w_{ik}$.
Particular examples of $w$ can be found in the literature.
For instance, the cases of a cold dust cloud \cite{polnarev}
and of a magnetized plasma \cite{macedo} are well established
and don't need to be explicitly given here. The contribution
to dispersion from the background curvature has also been
calculated for various space-times, see e.g. \cite{polnarev}.

In order to deal with spherical waves from a localized source it 
is appropriate to use a spherical coordinate system $(r, \theta, \phi)$,
such that: 

\begin{equation}
x^0 = c t, \quad x^1 = r, \quad x^2 = \theta, \quad x^3 =
\phi. \label{eq:3.5} 
\end{equation} 

For waves propagating in the radial direction we have to replace
solution (\ref{eq:3.2}) by the following wave solution

\begin{equation}
h_{ik} = \epsilon_{ik} \frac{a}{r} \exp (i q_0 x^0 + i q_1 x^1) 
 = \epsilon_{ik} \frac{a}{r} \exp (i q r - i \Omega t), \label{eq:3.6}
\end{equation} 
where $a$ is the new amplitude, $q = q_1$ and $\Omega = - q_0 c$. Because the
nondiagonal components $\eta^{ik}$, with $i \neq k$, are equal to
zero, we can easily 
transform the dispersion relation (\ref{eq:3.4}) into 

\begin{equation}
\left( \frac{\Omega^2}{c^2} - q^2 \right) = w (r, \theta, \phi),
\label{eq:3.7} 
\end{equation} 
where we have retained the possibility of a non-uniform distribution
of matter. We can see 
that the matter distribution can change the phase velocity of the
gravitational wave, according to 

\begin{equation}
v_f = \frac{\Omega}{q} = \sqrt{c^2 + \frac{w}{k^{2}}} \simeq c + c\frac{w}{2 q^2}
\label{eq:3.7} 
\end{equation} 

For the group velocity, we have

\begin{equation}
v = \frac{\partial \Omega}{\partial q} = \frac{q c^2}{\Omega} 
\left( 1 + \frac{1}{2 q} \frac{\partial w}{\partial q} \right) 
\simeq c \left( 1 + \frac{1}{2 q} \frac{\partial w}{\partial q} \right)
\sqrt{1 - wc^2/ \Omega^2}.  
\label{eq:3.8} 
\end{equation}

It is clear that, even if $w$ is independent of $q$, the presence of a
small amount of  
matter leads to a group velocity wave dispersion. It is known from
Nonlinear Optics that  
wave dispersion is an essential ingredient of self-phase modulation
\cite{luis}. It is the absence 
of dispersion that eventually explains the otherwise counter-intuitive
result that self-phase modulation  
is absent for plane gravitational waves propagating in empty flat
space-time \cite{servin}. 
The inclusion of matter is thus an essential ingredient of the present
study.    

\section{Nonlinear wave propagation}

We can now examine the possibility of a given radial wave, satisfying
the above linear dispersion relation, to interact with itself, due to
the nonlinear contributions contained in the term $R_{ik}^{(3)}$. 
The nonlinear contributions of $S_{ik}$ could equally be included,
but for simplicity they are neglected here. The existence of nonlinear
wave coupling 
implies that the wave amplitude $a$ in equation (\ref{eq:3.6}) can no
longer be a constant, and is replaced by a slowly varying function of
$r$ and $t$. 
This means that we now have

\begin{equation}
\partial_j h_{ik} = ( i q_j + \frac{1}{a} \partial_j a)
h_{ik}. \label{eq:4.1} 
\end{equation} 
So, we can write

\begin{equation}
\Box^2 h_{ik} \simeq [ - \eta^{jn} q_j q_n + i \eta^{jn} q_j \ln a ] h_{ik}. 
\label{eq:4.2} \end{equation}

Assuming that the above linear dispersion relation still holds, we can
cancel the first 
of these terms with the linear contribution from $S_{ik}$. For the
second term, we 
can write it as

\begin{equation}
i \eta^{jn} q_j \partial_n = - i q \partial_r - i \frac{\Omega}{c^2}
\partial_t  
= - i q (\partial_r + \frac{1}{v} \partial_t) \label{eq:4.3} 
\end{equation}
where $v = c^2 / v_f$ is the group velocity.

In order to establish the nonlinear equation for the slowly varying amplitude
$a$ we now use an approximate expression for $R_{ik}^{(3)}$, where only the
terms oscillating at the frequency of the wave $\Omega$ are retained

\begin{equation}
R_{ik}^{(3)} \simeq \frac{1}{2} \left( \frac{\Omega^2}{c^2} - q^2 \right) 
\frac{| a |^2}{r^2} \frac{a}{r} \exp(i q r - i \Omega t). 
\label{eq:4.4} 
\end{equation}

In deriving this expression we have added to the solution (\ref{eq:3.6}) 
its complex conjugate, in order to adequately describe a real wavepacket.
For a given line of sigth between the source  at $r \simeq 0$ and the 
eventual observer at a finite distance $r$, we can use equation
(\ref{eq:3.7}) with 
fixed values of $\theta$ and $\phi$. Replacing in the nonlinear wave
equation (\ref{eq:2.3}) we obtain

\begin{equation}
i q (\partial_r + \frac{1}{v} \partial_t) a = w (r) \frac{| a |^2}{r^3} a
\label{eq:4.5} 
\end{equation}

This nonlinear equation for the slow wave amplitude clearly shows
 that the nonlinear effects disappear in the absence of matter, $w (r) = 0$,
as noticed previously \cite{servin}. Let us make a variable transformation
from the pair $(r, t)$ to $(z, \tau)$, where we define $z = r -  v t$ and 
$\tau = t$. We have then $\partial_r = \partial_z$ and 
$\partial_t = \partial_\tau - v \partial_z$. Replacing this in the above
equation, we get

\begin{equation}
\partial_\tau a = - i \frac{w(z, \tau)}{q} v \frac{| a |^2}{r^2(z, \tau)} a 
\label{eq:4.6} 
\end{equation}

This equation is satisfied by a solution of the from

\begin{equation}
a (z, \tau) = a (z) \exp[ i \phi(z, \tau)] \label{eq:4,7} 
\end{equation}
with the phase function determined by

\begin{equation}  
\phi(z, \tau) = \phi_0 - \int^\tau \frac{w(z, \tau')}{q} v \frac{| a
  |^2}{r^2(z, \tau')} d \tau' 
\label{eq:4.8} \end{equation}

This solution represents a gravitational wavepacket propagating
spherically with a nearly  
constant envelope $a (z)$ and a variable nonlinear phase. The wave
frequency shift 
$\Delta \Omega$ will be given by the derivative of this phase with
respect to the 
time variable $t$ \cite{alfano}. Neglecting the small variation of the
distance $r (z, \tau)$ and the matter  
distribution $w (z, \tau)$ inside the wavepacket envelope, this means
that $\Delta \Omega$ 
will be essentially due to the variation of the energy distribution $|
a (z)|^2$ with respect 
to time. But this envelope is only a function of $z = r - v t$ and we can use
 
\begin{equation}
\partial_t | a (z) |^2 = - v \partial_z | a (z)|^2 \label{eq:4.9}
\end{equation} 

We can then state that

\begin{equation}
\Delta \Omega (\tau) = - v \partial_z \phi (z, \tau).
\label{eq:4.10} 
\end{equation}

Noting that, for  short wavepackets, the variation of the matter
dispersion term  
$w (z, \tau)$ and distance with respect to the source $r (z, \tau)$ can be 
negligible, we replace them in the expression of the phase 
by their central values $w (\tau) = w (z=0, \tau)$ and $r (\tau) = r
(z = 0, \tau)$. 
This leads to the following expression of the frequency shift
occurring inside the  
wavepacket envelope

\begin{equation}
\Delta \Omega (\tau) = \int^\tau \frac{w(\tau')}{q}
\frac{v^2}{r^2(\tau')} \partial_z | a (z) |^2 d \tau' 
 \label{eq:4.11} 
\end{equation}

Here we notice that the distance travelled by the wavepacket can be written as
$r (\tau) = \int^\tau v(\tau') d \tau' \simeq c \tau$. Neglecting the
possible slow change on the 
shape of the envelope over distance, we can finally write the above
expression as 

\begin{equation}
\Delta \Omega (\tau) \simeq \frac{1}{q} \partial_z | a (z) |^2
\int_0^\tau \frac{w (\tau')}{\tau'^2} d \tau' 
\label{eq:4.12} 
\end{equation}

In order to understand the physical meaning of this result, let us consider
the simple case of a uniform distribution of matter along the entire
line of sight. We can use 
$w (\tau) \simeq w_0 = const.$, and get for the frequency shift, after
a distance $r \simeq c \tau$ 
travelled by the wakepacket

\begin{equation}
\Delta \Omega (r) \simeq - \frac{c}{q} w_0 \partial_z | A (z)|^2 r, 
\label{eq:4.13} \end{equation}
where we have used the local spherical wave amplitude $A (z) = a (z) / r$, 
observed at a distance $r$ from the source. This  
linear dependence of the frequency shift with time,
or with the travelled distance, was found previously for 
linear propagation \cite{mend02} and is well known from Nonlinear
Optics \cite{luis}. 
Here, however, the frequency shift is proportional to the square of the 
local amplitude, which means that this effect can only the significant if it
occurs over short distances, not far away from the emitter. This 
feature is specific of spherical waves propagating in uniform media.

Another interesting case is that of a non-uniform matter distribution
where the wavepacket propagates accross a succession of $N$ localized
clouds, at distances $r_i$ (with $i = 1... N$) from the source, and widths
$\Delta r_i \ll r_i$. We can then transform equation (\ref{eq:4.12}) in

\begin{equation}
\Delta \Omega (r) \simeq - \frac{c}{q} \sum_i w_i \partial_z |A_i (z)
|^2 \Delta r_i 
\label{eq:4.14} \end{equation}
where the local envelope amplitudes are determined by $A_i (z) = a (z) / r_i$.
Again, the strongest contribution to the total frequency shift will
result from the 
clouds located nearest to the source, supposing that they all have similar
matter densities.

Let us assume that the source emits a gravitational wavepacket with $n$ cycles.
Its width with be $\delta z \simeq 2 \pi n / q$. And the maximum
frequency shift 
associated with the closest cloud ($i = 1$) will be of the order of

\begin{equation}
\Delta \Omega_{max} \simeq \frac{c}{q} w_1 \frac{|A_1 (z =
  0)|^2}{\delta z} \Delta r_1 
\label{eq:4.15} \end{equation}

We can also write $w_1 = (\Omega / c)^2 \alpha$, where $\alpha \ll 1$
is a small 
dimensionless factor. Noting that $\Omega \simeq q c$, this allows us
to establish 
the necessary condition for a large frequency shift, leading to a significant
spectral energy dilution, as $\Delta \Omega \geq \Omega$, or equivalently

\begin{equation}
\alpha | A_1|^2 \frac{\Delta r_1}{\delta z} = \frac{\alpha}{2 \pi n} | A_1 |^2 
(q \Delta r_1) \geq 1 \label{eq:4.16} 
\end{equation}
 
Notice here that $(q \Delta r_1/ 2 \pi)$ is the number of wavelengths
over the cloud width.  
This can be a very large number. For instance, for $\Omega = 10^4$ (wave in the
KHz range) and a width of $\Delta r_1 = 10^{-1} parsec$, we get $(q
\Delta r_1) = 10^{13}$. 
This means that, for a short pulse $(n < 10)$, dense enough cloud
$(\alpha > 10^{-4})$ and 
close enough source $(|A_1| >10^{-4})$, the above criterium could be satisfied.
This confirms, in more solid grounds, the suggestion previously made
\cite{mend02} 
that self-phase modulation could eventually take place.

Some further discussion on the size of the effect of self-phase
modulation may be in place. In principle, close to the source the
self-phase modulation 
could be stronger than the value presented above, not only because
of larger gravitational wave amplitude, but mainly because of the significant 
angular dependence of the metric. This can be seen as follows. Observe 
that terms proportional
to $\Omega^2 - c^2q^2$ in Eq.~(\ref{eq:4.4}) implicitly contains an
angular dependence, due to the fact that for spherical waves $q =
q(r)$ is found from the spherical Bessel functions. Separating
variables in spherical coordinates, it can be inferred that $\Omega^2
- c^2q^2 \sim \ell(\ell -1)c^2/r^2$, $\ell$ being the mode number of
the associated Legendre polynomial. Thus, for high mode numbers, the
nonlinear modification may increase by a substantial amount. 

\section{Conclusions}

Nonlinear wave propagation of spherical gravitational waves was considered in
this work. The possible occurrence of self-phase modulation was 
discussed. The case of short wavepackets emitted from a point source
in flat space-time 
was examined, where a tenuous distribution of matter was retained
in order to guarantee linear wave dispersion, which is a necessary
condition for self-phase modulation to occur. 
An explicit criterium for a significant spectral energy dilution due
to self-phase modulation was established. It leads to the
conclusion that the occurrence of self-phase modulation due
to matter distribution very close to the gravitational wave source
is plausible. 

In contrast with what could occur with plane wave propagation,
for spherical waves the contributions of phase modulation over
distance decay very 
rapidly with distance from the source, due to wave amplitude
decrease. For this reason, self-phase modulation is dominantly occurring
very close to the emitter. Notice however that the region where this effect
takes place is not necessarily the region where it can be observed, because
the expanded wave spectrum will then propagate far away without further
changes.

The efficiency of the self-phase modulation process is directly dependent on
wave dispersion, which is a consequence of matter distribution. Curvature of
space-time would also contribute to wave dispersion and would enhance the 
process. If, instead of spherical wave emission we have some kind of
directionality, the wave amplitude decay will be smaller and phase modulation
will also increase. Another source of nonlinearity is the energy-momentum
tensor, or the matter distribution itself, which was not retained here.  
Space-time curvature, directionality effects and energy-momentum
nonlinearities  
will eventually lead to more favorable criteria for the occurrence of
self-phase 
modulation of gravitational waves, and will be considered in a future work.


\begin{thebibliography}{10}
\bibitem{tenuous} Here, tenuous means that the background radius
of curvature should be long compared to the wavelength of the
gravitational wave.
\end{thebibliography}
\end{document}